\documentclass[12pt]{article}

\usepackage{natbib}

\title{Complex systems science and urban science: towards applications to sustainability trade-offs in territorial systems}
\date{}
\author{J. Raimbault$^{1,2,3,4,\ast}$ and D. Pumain$^4$\medskip\\
$^1$ LASTIG, Univ Gustave Eiffel, IGN-ENSG\\
$^2$ CASA, University College London\\
$^3$ UPS CNRS 3611 ISC-PIF\\
$^4$ UMR CNRS 8504 Géographie-cités\medskip\\
$\ast$ \texttt{juste.raimbault@ign.fr}
}

\begin{document}

\maketitle

\begin{abstract}
Urban systems are at the core of current sustainability concerns, and their study from a complexity perspective has a long history in several disciplines. We survey this literature and discuss future research directions relevant to sustainable planning, in particular the construction of integrative approaches. We finally illustrate this research program with the coupling of urban simulation models to explore trade-offs between sustainable development goals in systems of cities.
\end{abstract}

\newpage

\section{Introduction}

The main ecological and societal challenges of this early 21st century are tightly intertwined into complex problems, partly captured by the Sustainable Development Goals (SDGs) as they were put forward by the United Nations \citep{nations2015sustainable}. These imply contradictory objectives implemented by multiple stakeholders at various scales. Cities and more generally urban systems are a central aspect to tackle these, concentrating simultaneously many issues (congestion, emissions, increased economic activities) but also solutions (social and technological innovation, economies of scale) related to sustainable development. While many disciplines have extensively studied these urban sustainability questions (urban economics, urban geography, sustainability science, political science to give a few), the rise of an interdisciplinary urban science \citep{batty2021new}, inheriting from former strong links between complexity science and urban questions, appears as a relevant candidate to bring new answers to the sustainability conundrum. Indeed, when looking at complexity from a theoretical viewpoint as Edgar Morin put it throughout the whole transdisciplinary perspective of La Méthode \citep{morin1991methode}, accounting for the intrinsic complexity of systems (whatever the operational definition of complexity used, e.g. chaotic dynamics or the presence of emergence) cannot be dissociated from a multi-scale understanding of systems, and therefore is a knowledge that transcends traditional disciplines. From an empirical viewpoint, an integration of dimensions seems necessary to handle the SDGs, due to the numerous negative (trade-offs) or positive (co-benefits) interactions between these goals \citep{nilsson2018mapping}.

This contribution aims at giving an overview of this research perspective focusing on complexity and urban systems. More precisely, we survey the existing links between complex systems approaches and urban science, suggest some paths forward for the application of such integrative approaches to the planning of sustainability, and illustrate this with a case study of urban system dynamics models applied to the search of trade-offs between SDGs.

\section{Complexity and urban science}

We first give a broad overview of how scientific paradigms related to complexity have been applied to the study of urban systems in the literature. This is far from an exhaustive review, as the goal is rather to highlight the diversity of methods and the overarching complexity of urban systems.

The link between complexity approaches and the study of urban systems has historically always been strong, starting already with the precursors. The systems dynamics modelling technique, which was developed in the early 70s through the transfer of concepts from cybernetics \citep{schwaninger2008system}, found its most notable application with the Meadows report on the limits of growth \citep{meadows1974dynamics}, but was also an important entry for enquiries on urban complexity at larger scales: \cite{forrester1970urban} work on urban dynamics was among the first to propose a holistic simulation approach of such systems. This modelling technique later diffused into quantitative geography, with applications at regional scales \citep{chamussy1984dynamique}, in some cases related to decision-making stakeholders.

Another stream of research with fruitful applications to the study of cities was fractals, in particular to the understanding and quantification of urban form \citep{batty1994fractal}. A fractal nature of the urban fabric and different fractal dimensions have implications for various urban phenomena, including for example social dynamics, urban climate, energy efficiency, access to amenities. This approach is still active nowadays, in the theoretical \citep{chen2018logistic}, empirical \citep{salat2018uncovering} or applied fields \citep{frankhauser2018integrated}. The physics of dissipative structures explored in the 80s following Prigogine also found rapid applications to the modelling of urban systems. According to \cite{pumain1984vers}, an intra-urban dynamical model proposed by P. Allen is relevant for planning application, while a comparable model by A. Wilson has a more robust theoretical basis while being more difficult to apply. Relatedly, the foundational work by \cite{wilson1971family} on spatial interaction modelling is a direct application of entropy maximisation concepts imported from statistical physics.
Around the year 2000 flourished different fields related to complexity. The study of complex networks witnessed a theoretical renewal associated with new empirical observation, data and models with rapid applications to the study of urban networks \citep{derudder2018uncovering,neal2021handbook} or urban street networks \citep{jiang2004topological} - the later being already investigated with a different theoretical background for a long time through the space syntax approach for example \citep{hillier1976space}.

The study of urban scaling laws somehow got a lot of attention at around the same time. They express different urban indicators as scaling with city size, either sub- or super-linearly through a power-law \citep{pumain2004scaling}. The particular case of city size power law had already been known and studied since at least the beginning of the 20th century with precursors such as G. Zipf. Explaining the striking regularity of such power laws across urban systems is still an open question \citep{ribeiro2021mathematical}, for which inter-urban innovation dynamics are a possible explanation for example \citep{pumain2006evolutionary}.

This last example relates to the study of urban evolution, in the sense of a geographical theory accounting for the complex and adaptive nature of urban systems introduced by \cite{pumain1997pour}, and which can be interpreted as an extension of social and cultural evolution \cite{raimbault2020model}. This stream of research has been fruitful for the development of complexity approaches in urban studies, with among the first agent-based models applied to a geographical system \citep{sanders1997simpop}, and more recently a collection of urban systems simulation models applied to urban systems worldwide and an associated set of tools and methods to explore and validate spatial simulation models \citep{pumain2017urban,reuillon2013openmole}.
The use of Cellular Automatons models to simulate the growth of urban form also quickly developed at the same time \citep{batty1997cellular}. Many operational land-use change models are now based on this paradigm. The question of urban morphogenesis, in particular how simple processes can be complementary to simulate urban growth \citep{raimbault2020comparison} and the link between urban form and function, remains rather open. The transfer of complexity paradigms originating in biological sciences, such as the field of artificial life, finds relevant applications in the study of urban systems. The study of co-evolution in urban systems is core to \cite{pumain2018evolutionary} theory of urban system, and was recently modelled in the case of transportation networks and territories by \cite{raimbault2018caracterisation}.

These examples of complexity approaches of urban systems are not exhaustive - other complexity related fields such as participatory modelling, game theory, chaos, statistical physics, microsimulation, artificial life, artificial intelligence, etc., have found application in urban settings; see e.g. \citep{raimbault2020cities} for a literature mapping in the case of artificial life). This however illustrates the productive exchanges between urban science and complexity in history and in many contemporary fields still very active.

\section{Perspectives towards sustainable planning}

Within this broad framework of urban complexity, we can sketch some research directions that we estimate crucial to address the current challenge of sustainable transitions. More particularly, we postulate that the construction of integrated approaches will be fruitful for sustainable planning. This position was developed with more details by \cite{raimbault2021integrating}. Although a proper definition of ``integration'' still lacks, we consider the coupling of heterogeneous simulation models as a medium to couple and thus integrate perspectives. Following the Complex Systems roadmap \citep{2009arXiv0907.2221B}, integration can either be horizontal (transversal research questions spanning all types of complex systems) or vertical (construction of multi-scalar integrated disciplines). In terms of urban modelling, this translates in the case of horizontal integration through multi-modelling \citep{cottineau2015modular}, model benchmarking \citep{raimbault2020comparison} and model coupling. This horizontal integration is necessary to capture the complementary dimensions of urban systems, and the potentially contradictory SDGs. Vertical integration relates to the construction of multi-scalar models accounting for both top-down and bottom-up feedbacks between scales, which is still an open issue but remains essential for the design of policies suited for each territory \citep{rozenblat2018conclusion}. The three typical scales to be accounted for are the micro scale of intra-urban processes, the meso scale of the urban area, and the macro scale of the system of cities \citep{pumain2018evolutionary}.

Within this context, reflexivity is key to ensure a consistent horizontal integration, and the development of associated tools providing literature mapping and corpus exploration is a component of this research program \citep{raimbault2021empowering}. This furthermore requires a full practice of open science, to understand and validate the components integrated. This validation must furthermore be achieved using dedicated methods and tools. The OpenMOLE platform for model exploration \citep{reuillon2013openmole} provides a seamless framework to embed models, couple them through the workflow system, and apply state-of-the-art sensitivity analysis and validation methods using high performance computing.

To summarise, we claim that sustainable planning taking into account contradictory SDGs can be achieved using integrated approaches. A first step of this long-term research objective relies on the coupling of heterogeneous urban simulation models and the construction of multi-scale urban models, both facilitated by innovative model validation methods and tools. The question of the actual transfer of results to policies remains an open question at this stage, but for which several suggestions can be done, including e.g. participatory modelling and the interactivity of models (see \citep{raimbault2021integrating}).

In the remainder of this contribution, we illustrate the application of this framework to the specific case of trade-offs between SDGs in synthetic systems of cities, using systems of cities simulation models.

\section{Trade-offs between SDGs in systems of cities}

\subsection{Urban dynamics and innovation diffusion: bi-objective trade-offs}

In \citep{raimbault2022trade}, an urban system dynamics model coupling population dynamics with the diffusion of innovation is applied to the search for bi-objective trade-offs in synthetic systems of cities. More precisely, the urban evolution model described by \citep{raimbault2020model} based on the innovation diffusion model by \citep{favaro2011gibrat} considers cities characterised by their population and an ``urban genome'' which consists in adoption shares of a given innovation. Iteratively, the model updates population through spatial interaction models, with an attractivity given by the level of innovation. In turn, innovations are diffused between cities using another spatial interaction model. Finally, new innovations emerge randomly in cities following a scaling law of population. The model is applied to synthetic systems of cities, and two SDGs are captured using proxies for transport emissions (spatial interaction flows) and for innovation. Using a genetic algorithm for optimisation, we obtain a Pareto front between these two objectives, confirming the existence of a trade-off in this setting. Varying the scaling exponent for new innovations, we find that a more balanced innovation system is always preferable rather than something highly centralised.

\subsection{Towards many-objective trade-offs}

The previous results remain highly stylised and within a reduced dimension objective space. Current work in progress aims at extending this work on different points, for which several difficulties however arise.

The first extension is to investigate whether trade-offs occur in practice. The investigation of empirical stylised facts on SDGs proxies compared across different urban systems would provide an insight, but remains limited by the lack of unified and comparable databases. As a comparison, constructing a consistent database for populations including 7 large urban systems worldwide necessitated a large effort including an ERC project and several PhD students \citep{pumain2015multilevel}. Extending this to multiple dimensions requires mapping existing databases, potentially undergoing some time-consuming collection work, and finally harmonising the data to ensure comparability.

A second and related point lies in the parametrisation of simulation models on real configurations rather than synthetic systems. In the particular case of urban dynamics and innovation diffusion, population data from \citep{pumain2015multilevel} has already been used to benchmark such macroscopic simulation models \cite{raimbault2020empowering}. The innovation data is however much more problematic. A standard entry is to use patent data as a proxy for innovation. Recent patent datasets have been geocoded in terms of inventor address, for example by \citep{de2019geocoding}. Historical geocoded patent data (for example before 1976 for the US Patent Office), which is needed for this model running on long time scales, is much more rare to find. An initiative to build such an open database for Europe is currently ongoing by \citep{bergeaud2021patentcity}. Then, either an extension of the innovation diffusion model to include multiple types of innovations (matrix genome instead of a vector genome), or a selection of typical technological classes that were empirically correlated with city attractivity, would be necessary. An alternative approach can rely on the semantic content of patents rather than their exogeneous classification \citep{bergeaud2017classifying} to better characterise the spatial diffusion of innovation and parametrise the model using the principal component across semantic dimensions.

Finally, extending the study of trade-offs across other SDGs - there are 17 distinct goals in total, with numerous subgoals and quantitative indicators - is an important research direction. The aforementioned model shares a common basis with other urban systems dynamics models \citep{pumain2017urban}, and thus can be coupled with these to include other dimensions. An economic exchange model described by \citep{cottineau2015modular} allows including the aspects of wealth and inequalities, while a co-evolution model between cities and transport networks introduced by \citep{raimbault2021modeling} accounts for infrastructure. These layers - which were separately benchmarked by \citep{raimbault2020empowering} - are strongly coupled into a multi-modelling framework. The resulting simulation model provides 5 proxy indicators for SDGs, and is being explored and optimised on these possibly conflicting dimensions using a many-objective genetic optimisation algorithm.

\section*{Conclusion}

Complexity approaches have always cultivated deep ties with studies of urban systems in different disciplines and more recently with the emerging urban science. We believe that the future of urban sustainability cannot be conceived without such a multidimensional and complex approach, and therefore suggest that one relevant research direction among others to achieve this is the construction of integrated approaches, in practice by coupling urban simulation models and building multi-scale urban models.

%\bibliographystyle{apalike}
%\bibliography{biblio}

\end{document}